\documentclass[twoside,twocolumn,9pt]{article}
\usepackage{extsizes}
\usepackage[super,sort&compress,comma]{natbib} 
\usepackage[version=3]{mhchem}
\usepackage[left=1.5cm, right=1.5cm, top=1.785cm, bottom=2.0cm]{geometry}
\usepackage{balance}
\usepackage{mathptmx}
\usepackage{sectsty}
\usepackage{graphicx} 
\usepackage{lastpage}
\usepackage[format=plain,justification=justified,singlelinecheck=false,font={stretch=1.125,small,sf},labelfont=bf,labelsep=space]{caption}
\usepackage{float}
\usepackage{fancyhdr}
\usepackage{fnpos}
\usepackage[english]{babel}
\addto{\captionsenglish}{%
  
}
\usepackage{array}
\usepackage{droidsans}
\usepackage{charter}
\usepackage[T1]{fontenc}
\usepackage[usenames,dvipsnames]{xcolor}
\usepackage{setspace}
\usepackage[compact]{titlesec}
\usepackage{hyperref}

\usepackage{epstopdf}
\usepackage[dvipsnames]{xcolor}
\usepackage[normalem]{ulem}
\usepackage{arydshln}
\usepackage{multirow}

\newcommand{\orange}[1]{\textcolor{black}{#1}}

\definecolor{cream}{RGB}{222,217,201}

\begin{document}

\pagestyle{fancy}
\thispagestyle{plain}
\fancypagestyle{plain}{
\renewcommand{\headrulewidth}{0pt}
}

\makeFNbottom
\makeatletter
\renewcommand\LARGE{\@setfontsize\LARGE{15pt}{17}}
\renewcommand\Large{\@setfontsize\Large{12pt}{14}}
\renewcommand\large{\@setfontsize\large{10pt}{12}}
\renewcommand\footnotesize{\@setfontsize\footnotesize{7pt}{10}}
\makeatother

\renewcommand{\thefootnote}{\fnsymbol{footnote}}
\renewcommand\footnoterule{\vspace*{1pt}%
\color{cream}\hrule width 3.5in height 0.4pt \color{black}\vspace*{5pt}} 
\setcounter{secnumdepth}{5}

\makeatletter 
\renewcommand\@biblabel[1]{#1}            
\renewcommand\@makefntext[1]%
{\noindent\makebox[0pt][r]{\@thefnmark\,}#1}
\makeatother 
\renewcommand{\figurename}{\small{Fig.}~}
\sectionfont{\sffamily\Large}
\subsectionfont{\normalsize}
\subsubsectionfont{\bf}
\setstretch{1.125} 
\setlength{\skip\footins}{0.8cm}
\setlength{\footnotesep}{0.25cm}
\setlength{\jot}{10pt}
\titlespacing*{\section}{0pt}{4pt}{4pt}
\titlespacing*{\subsection}{0pt}{15pt}{1pt}

\fancyfoot{}
\fancyhead{}
\renewcommand{\headrulewidth}{0pt} 
\renewcommand{\footrulewidth}{0pt}
\setlength{\arrayrulewidth}{1pt}
\setlength{\columnsep}{6.5mm}
\setlength\bibsep{1pt}

\makeatletter 
\newlength{\figrulesep} 
\setlength{\figrulesep}{0.5\textfloatsep} 

\newcommand{\topfigrule}{\vspace*{-1pt}%
\noindent{\color{cream}\rule[-\figrulesep]{\columnwidth}{1.5pt}} }

\newcommand{\botfigrule}{\vspace*{-2pt}%
\noindent{\color{cream}\rule[\figrulesep]{\columnwidth}{1.5pt}} }

\newcommand{\dblfigrule}{\vspace*{-1pt}%
\noindent{\color{cream}\rule[-\figrulesep]{\textwidth}{1.5pt}} }

\makeatother

\twocolumn[
  \begin{@twocolumnfalse}
{

}\par
\vspace{1em}
\sffamily
\begin{tabular}{m{4.5cm} p{13.5cm} }

& \noindent\LARGE{\textbf{\orange{Explaining the spread in measurement of PDMS elastic properties: influence of test method and curing protocol}}} \\
\vspace{0.3cm} & \vspace{0.3cm} \\

& \noindent\large{Hannah Varner,\textit{$^{a}$} and Tal Cohen\textit{$^{a,b,*}$}} \\

\includegraphics{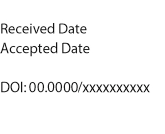} & \noindent\normalsize{\orange{Accuracy in the measurement of mechanical properties is essential for precision engineering and for the interrogation of composition-property relationships. Conventional methods of mechanical testing, such as uniaxial tension, compression, and nanoindentation,  provide highly repeatable and reliable results for stiff materials, for which they were originally developed. However, when applied to the characterization of soft and biological materials, the same cannot be said, and the spread of reported properties of similar materials is vast. 
Polydimethylsiloxane (PDMS), commonly obtained from Dow as SYLGARD~184, is a ubiquitous such material, which has been integral to the rapid development of biocompatible microfluidic devices and flexible electronics in recent decades.  However, reported shear moduli of this material range over 2 orders of magnitude for similar chemical compositions. Taking advantage of the increased mechanical scrutiny afforded to SYLGARD~184 in recent years, we combine both published and new experimental  data obtained using 9 mechanical test methods. A statistical analysis then elucidates the significant bias induced by the test method itself, and  distinguishes this bias from  the influence of curing protocols on the mechanical properties. 
The goal of this work is thus two-fold: (i) it provides a quantitative understanding of the different factors that influence reported properties of this particular material,  and (ii) it serves as a cautionary tale. As researchers in the field of mechanics strive to quantify the properties of increasingly complex soft and biological materials, converging on a standardized measurement of PDMS is a necessary first step.}}

\end{tabular}

\end{@twocolumnfalse} \vspace{0.6cm}
]


\renewcommand*\rmdefault{bch}\normalfont\upshape
\rmfamily
\section*{}
\vspace{-1cm}


\footnotetext{\textit{$^{a}$~Department of Mechanical Engineering. Massachusetts Institute of Technology. Cambridge, MA.}}
\footnotetext{\textit{$^{b}$Department of Civil and Environmental Engineering. Massachusetts Institute of Technology. Cambridge, MA.}}
\footnotetext{\textit{$^{*}$~Corresponding author.  E-mail: \href{mailto:talco@mit.edu}{talco@mit.edu}. } 77 Massachusetts Avenue. Cambridge, MA 02139.}

\footnotetext{\dag~Electronic Supplementary Information (ESI) available: Details of original data published here for the first time, Summary of all articles included across the meta-analysis, including interconnection mapping of citations. Additional details of regression analysis. See DOI: XXXXXX/}


\section{Introduction}
\label{sec:intro}

\orange{Modern engineering of material systems relies on the testing and quantification of a material's mechanical response.  The tension, compression, indentation and impact testing methods that most engineers first encounter are all standard for metal samples but may not be appropriate for soft materials, which are significantly more challenging to test. Nonetheless, the use of soft materials is becoming increasingly prevalent in engineering, as researchers try to better understand and interface with the human body, and develop smaller and more compliant material structures and devices.  Comparing measurements from a range of  techniques, several studies\cite{ji2022comparison, malone2018mechanical, varner2023elasticity} report  wide discrepancies in measured stiffness of soft biological materials. It is notable that despite well documented variations in biological tissue, these discrepancies are often attributed primarily to the testing method and not to material variations.}

\orange{Even commercially available synthetic materials are hard to test. One example is Polydimethylsiloxane (PDMS) where reported elastic moduli range by 2 orders of magnitude for similar compositions, as shown herein.}
PDMS is a two-part material system and the stiffness of the resulting silicone polymer can be easily tuned to suit the needs of an application. PDMS will cure at room temperature, or can have the curing accelerated with an oven cure. 
It is widely used in microfluidics \cite{mcdonald2002poly, xia1998soft, raj2020pdms}, medical devices, and electronics \cite{Lotters_1997, wu2015inkjet, wu2010flexible}. In engineering mechanics, it is a common  \orange{surrogate material for method development in the measurement of  elastic, adhesion, and fracture response \cite{zhang2021relationship, raayai2019capturing, kothari2020controlled, patel2019investigation, peng2011multi}, and as a component in the development of tough materials that can undergo large deformations before failure\cite{wang2019stretchable, jung2012cnt, ariati2021polydimethylsiloxane, niu2007characterizing, chen2015thermal, kowol2022strain, jiao2021composites}}. 
The flexibility of composition and cure is particularly useful in the context of biological research where PDMS has been proposed as a scaffold for \orange{cell growth \cite{brown2005evaluation,kim2007establishment, tanaka2007micro,fuard2008optimization}, as a biological membrane mimic in microfluidics \cite{huh2010reconstituting}, and in biohybrid devices \cite {feinberg2007muscular, Nawroth2012, chen2018polydimethylsiloxane}}. \orange{Given its widespread use in biology, numerous authors have worked to modify  the adhesion, wetablity, absorptivity, dielectric properties, and mechanical characteristics of PDMS,   with comprehensive reviews  provided by Wolf et al. \cite{wolf2018pdms}, Zhou et al.\cite{zhou2010recent}, Abbasi et al. \cite{abbasi2001modification}, Teixeira et al. \cite{teixeira2021polydimethylsiloxane} and  Murphy et al. \cite{murphy2020tailoring}. However, far less attention has been paid to systematically understanding and measuring the bulk properties of PDMS (or, often, the {commercially} branded product Dow SYLGARD~184 \cite{dowsylgard184}). The same fabrication flexibility and  wide availability that makes commercial SYLGARD 184 applicable across a range of fields also {highlights} a critical problem of mechanical reproducibility. }

\orange{The primary goal of this work is to elucidate  the bias induced by the testing method in determining the elastic response of soft materials. To this end, we choose SYLGARD 184 as a test case, taking advantage of the breadth of available data in the literature spanning a range of test methods, and combined with new Volume Controlled Cavity Expansion (VCCE) data, reported here. The secondary goal of this work is, thus, the interrogation of SYLGARD 184 mechanical properties. 
While frequent users of SYLGARD 184 often expect minor inconsistencies in the mechanical properties due to changes from the manufacturer\footnote{\orange{The exact composition of SYLGARD 184 is proprietary. Reading the product literature and safety data sheets provides the following insights~\cite{wolf2018pdms,heinrichs2018chemically,palchesko2012development,schweitzer2019determination}: The base contains a majority 
dimethylvinyl-terminated dimethyl siloxane ($>$60\%), with a significant portion of silica fillers (30-60\% dimethylvinylated and
trimethylated silica), as well as 1-5\% tetra(trimethylsiloxy) silane and small amounts of ethylbenzene, xylene and a platinum catalyst (<1\% each). The curing agent is a majority the cross linking agent dimethyl methylhydrogen siloxane (40-70\%), additional methylated silica and siloxanes, and 1-5\% the inhibitor
tetramethyl tetravinyl cyclotetrasiloxane}}, we hypothesize instate that measurement discrepancies can be mainly attributed to bias imparted by: (1) different mechanical testing methods, and (2)  wide variation in the cure conditions used for similar mixing ratios.} 

``Round Robin'' studies have been used to compare test methodologies and increase the ``reliability and repeatability'' of measurements in ceramics \cite{kubler1997fracture, quinn1994fracture}, composites \cite{davies1992round}, adhesives \cite{blackman2003measuring}, concrete \cite{van1997strain}, steel \cite{ruiz2020round}, and many other materials and industries.
In their Versailles Advanced Materials and Standards report, Kubler et al. \cite{kubler1997fracture} articulate their motivation for conducting the study to determine if the Single-edge-V-Notched beam method for measuring fracture toughness is ``user-friendly, reliable and ... comparable to other recognised methods''. A similar effort in the context of PDMS elasticity characterization is, in the view of the authors, overdue. To this end, we embark on this study in an effort to understand the magnitude of the problem at hand. We present the scale of variability observed across literature and in our own data, and hope to motivate the community to converge on more consistent  preparation, and testing methodologies to improve repeatability for PDMS and across the field of soft material mechanics.

This paper is organized as follows: we begin by describing how studies were chosen for inclusion in this meta-analysis~(Sec. \ref{sec:include}) and provide a brief description of the mechanical testing methods used across the included studies~(Sec. \ref{sec:tests}). We then present the results of our analysis~(Sec. \ref{sec:results}) and include new data for moduli obtained via VCCE (detailed in Sec. 2 of the ESI, Table ESI.T1). Finally, we conclude with remarks and cautions on stiffnesses reported in literature for PDMS~(Sec. \ref{sec:conclusion}).

\section{Methods}
\label{sec:methods}

\subsection{Data gathering}
\label{sec:include}
Articles surveyed for this review were found through key word searches and references lists of included articles. Keywords included ``elasticity'' or ``modulus'' or ``stiffness'' and ``PDMS'' or ``polydimethylsiloxane'' or ``SYLGARD~184'', as well as related derivatives and searches were performed across Google Scholar and Web of Science from January - August 2023. Studies included in the results presented here were those with defined a) mix ratios for curing agent and base, b) cure time, and c) cure temperature. We restrict this study to Dow's SYLGARD~184. Hence, studies that used other commercial formulations such as GE RTV PDMS silicones were excluded (such as by Liu et al.\cite{liu2005polymer} and some results reported by Schneider et al. \cite{schneider2008mechanical}). Note that one of the most commonly cited studies on PDMS moduli (see Fig. ESI.F1 in ESI.1), L\"{o}tters \cite{Lotters_1997} has not been included as the authors used ABCR PS851 PDMS.  

All included studies are summarized in Table ESI.T2 of Sec. ESI 3. Studies with parameters listed as ``variable'' intentionally survey a large range of values and are therefore not listed in the table (though included where appropriate in corresponding figures). \orange{New data collected by the authors and included in Table ESI.T2, is described in Sec. ESI.2 with fabrication and measurement detailed in ESI.T1.} 

Dow recommends a mixing ratio of 10 parts base to 1 part curing agent (crosslinker) on a weight to weight basis; we will abbreviate this as 10:1 or more generally w$_B$:w$_C$. Where the mix ratio was instead reported as a percent, we have converted to the \textit{X parts base:1 part curing agent} format.
All stiffnesses are reported here as shear modulus, $\mu$. Where  authors initially reported the stiffness as an elastic or Young's Modulus, $E$, incompressibility is assumed with a Poisson's ratio $\nu=0.5$ \cite{armani1999re} and $\mu=E/(2(1+\nu))=3E$. For studies investigating viscoelastic material parameters or high strain rate responses, the quasistatic modulus is reported in the summary charts and where only an instantaneous modulus is reported, we denote it as $\mu_0$. For studies reporting a cure at room temperature, 25~$^\circ$C was used in plotting and comparison tables.

\begin{figure*}[h]
    \centering
    \includegraphics[width=171mm]{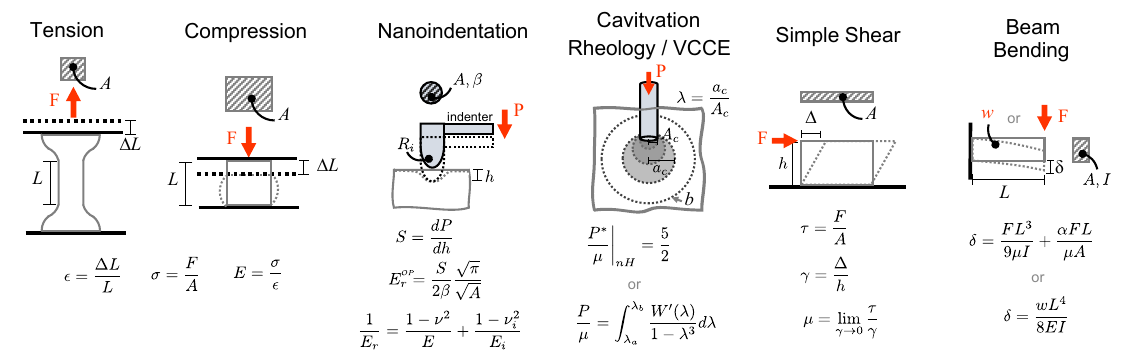}
    \caption{Common types of testing used to determine shear or elastic modulus, $\mu$ and $E$ for PDMS. Variable definitions are included in the text. } 
    \label{fig:test descriptions}
\end{figure*}

\subsection{Test methodologies}
\label{sec:tests}
Authors have used a diverse range of test methods to establish the stiffness of PDMS depending on the curing agent ratio in question and goal of the study. The fundamental assumptions and geometries of the different methods are discussed below and depicted in Fig. \ref{fig:test descriptions}. 

\textit{Tension}: Tension testing was undertaken using a commercial tension test machine in all cases except Brown et al. \cite{brown2005evaluation} who use a fixed applied weight, Upadhyay et al.\cite{upadhyay2021validated} who perform dynamic testing with a modified split Hopkinson bar, and  Fuard et al. \cite{fuard2008optimization} who use a custom stretching tool and load cell. Tracking the force $F$, and sample extension $\Delta L$, from initial length $L$, the material stiffness is calculated as $\sigma=E\epsilon$ in the linear regime where $\sigma=F/A$ and $\epsilon=\Delta L/L$ are the stress and strain, respectively, for a sample cross sectional area,~$A$.  Clamp displacement is used  for tracking average sample extension in~\cite{khanafer2009effects, kim2011measurement,liu2009influences, johnston2014mechanical, wang2019stretchable, schneider2008mechanical}, while Upadhyay et al.~\cite{upadhyay2021validated} use particle tracking and digital image correlation, Mills et al.~\cite{mills2008mechanical}, Seghir and Arscott~\cite{seghir2015extended} and Fuard et al.~\cite{fuard2008optimization} use optical image capture and post process with a known calibration, and Brown et al.~\cite{brown2005evaluation} use direct measurement. The ASTM Standard D412 for a ``Standard Test Methods for Vulcanized Rubber and Thermoplastic Elastomers—Tension''~\cite{astm412} includes a set of defined sample geometries, ramp rates, and test conditions and was used by \cite{johnston2014mechanical,khanafer2009effects,mouvcka2021mechanical,liu2009influences,mata2005characterization} who all focus on only a 10:1 base to curing agent ratio.

\textit{Compression}: Unconfined compression (using $\sigma=E\epsilon$ to determine the elastic stiffness in the linear regime as with tension testing) was employed by Carrillo et al.~\cite{carrillo2005nanoindentation} to compare against nanoindentation, as well as by Wang et al.~\cite{wang2014crosslinking} in their comparison of moduli across curing agent ratios.

\textit{Nanoindentation}: 
With an indenter tip possessing a projected contact area $A$, and geometry described by $\beta$, the  modulus of a material near a free surface can be determined from the slope of the pressure-displacement curve $S=dP/dh$ of an indentation. Calculating the reduced modulus $E_r$ based on the indenter geometry, then combining with the indenter modulus $E_i$ and Poisson's ratio $\nu_i$ returns the sample modulus, $E$.
Commercially available hardware was used by most authors, however, Mata et al.\cite{mata2005characterization} were the only authors to use the standard analysis method included with their tool, while all other authors reported specifics of their analysis: Patel et al.\cite{patel2019investigation} used the Oliver Pharr method (not accounting for adhesion) to determine $E_r^{OP}$. To avoid the overestimate of modulus that is typical in polymers when using Oliver Pharr, Cheng et al.\cite{cheng2011characterization} instead used a Hertzian contact model, whereby 
\[
   E_H= \sqrt{\frac{S^3(1-\nu^2)^2}{6R_iP_{max}}} 
\]
for a spherical indenter with radius of curvature $R_i$ and $\nu$ is assumed to be 0.5 for PDMS. Multiple authors acknowledge the significance of adhesion in nanoindentation measurements and employ the Johnson, Kendall, Roberts (JKR) model instead of Hertzian contact \cite{carrillo2005nanoindentation, cao2005nanoindentation, peng2011multi}, or in addition to it\cite{gupta2007adhesive, liao2010hybrid}. However, authors disagreed whether the Hertzian contact model returned a higher moduli than the JKR model \cite{carrillo2005nanoindentation} or a lower moduli \cite{peng2011multi}.

\textit{Volume Controlled Cavity Expansion (VCCE)}: Controlled injection of an incompressible working fluid at the tip of an injection needle is used to expand a spherical cavity in a sample~\cite{raayai2019volume,raayai2019capturing}.  The resisting pressure in the cavity, $P$, is simultaneously measured to obtain a nonlinear pressure-volume relationship, or equivalently $P=P(a_c)$, where $a_c$ is the effective cavity radius. Comparison with theoretical prediction of elastic, spherically symmetric, cavity expansion  in  incompressible hyperelastic materials, namely 
\begin{equation}\label{eq:vcce}
P(a_c)=\int_{a_c/A_c}^{1}\frac{W'(\lambda)}{1-\lambda^3}d\lambda,
\end{equation}
then allows to determine  material parameters. Where $\lambda$ is the circumferential stretch which varies from $a_c/A_c$ at the cavity wall to $1$ in the remote field,  $W(\lambda)$ is the corresponding elastic potential energy,  and $A_c$ is the initial defect size, which needs to be determined. This method can also be applied to probe viscoelastic  properties  \cite{chockalingam2021probing}.

Li et al.\cite{li2024cylindrical} use similar assumptions in a cylindrical geometry, finding a very good match between analytical and experimental results. 

\textit{Cavitation Rheology (CR)}: Alternately, a second type of cavity expansion, cavitation rheology, aims to identify the peak value of the resisting pressure, $P^*$, and assumes that it is well represented by the theoretical cavitation limit which is then used  to determine the material parameters. In contrast to VCCE, this approach  does not require control of the injected volume (it typically uses gas as an injection medium).  Milner \cite{milner2021multi, milner2020localized} and Yang et al. \cite{yang2019hydraulic} assume a neo-Hookean material model\footnote{This result can be obtained directly from Eq. \eqref{eq:vcce} at the limit $\lambda\to\infty$, with a neo-Hookean form of $W(\lambda)$, as shown in \cite{raayai2019volume}.}, which implies $P^*= \frac{5}{2}\mu_{nH}$. 

\textit{Simple Shear}: Both Upadhyay et al.\cite{upadhyay2019quasi} and Nunes \cite{nunes2011mechanical} assume linear (small strain) deformations in simple shear. An applied shear force, $F$, induces a displacement, $\Delta$, and corresponding shear strain $\gamma=\Delta/h$. The shear  modulus is then estimated as $\mu=\tau/{\gamma}$, where the shear stress is $\tau=F/A$ and $h$ and $A$ are sample height and area shown in Fig. \ref{fig:test descriptions}. 

\textit{Beam Bending}:
This method relies on connecting the displacement of a beam with know geometry to an applied load. Armani et al. \cite{armani1999re} measured tip deflection, $\delta$, of a length $L$ cantilever beam that was displaced under its own weight ($w$ per unit length) and calculated modulus using $\delta=\frac{w L^4}{8EI}$ with $I$ describing the second moment of area. By contrast, Du et al.\cite{du2010extension} considered a point load at the tip of a low aspect ratio beam and  added a Timoshenko term $\frac{\alpha  FL}{EA}$ with the shear coefficient $\alpha$.

\textit{Membrane Deflection}: A unique measurement method by Thangawng et al. \cite{thangawng2007ultra}, uses optical profilometry to measure the membrane deflection of a nanometer thickness layer of spin coated PDMS on a silicone wafer. By pressurizing a circular portion of the membrane with radius $r$ and thickness $t$, the authors were able to connect deflection, $\delta$, and material modulus as 
\[\frac{P}{\delta}=\frac{C_1 t}{r^2}  \sigma_0 + \frac{C_2 f(\nu) t}{r^4}\frac{E}{1-\nu}\delta^2\]
using geometric coefficients $C_1, C_2$ and $f(\nu)$ and residual stress $\sigma_0$, and again assuming $\nu=0.5$. 

\subsection{Correlation analysis}
\label{sec:correlation}
Compiled data is visualized to observe trends and the monotonicity of relationships is quantified using Spearman rank\cite{spearman1961proof} correlation coefficient, $\rho$. 
The Spearman correlation \orange{presents the null hypothesis that there is no correlation between an input (predictor) variable and the output, and returns $-1<\rho<1$ indicating whether a monotonic correlation exists, and whether it is negative or positive, respectively. A significance p$<0.05$ indicates confidence in rejecting the null hypothesis of the data originating from an uncorrelated dataset (more precisely, from a dataset that is less well correlated than indicated by the magnitude of $\rho$). Spearman correlation is used given the non-normal distribution of all predictor variables, and the desire to not impose a linear relationship between the inputs and stiffness \cite{everitt2010cambridge}. Further discussion of the statistical distribution of the predictor variables is given in Sec. 4 of the ESI. }

\orange{The correlation} is executed via MATLAB 2023 and is applied between preparation conditions (cure temperature, time) and within subgroups of individual test methods in order to isolate \orange{where monotonicity might hold internal to a test method, but not within the entire population}. While $\rho$ provides indication of how \textit{monotonic} the relation between two inputs is ($0.3\leq|\rho|<0.6$: Fair, $0.6\leq|\rho|<0.8$: Moderate, $0.8\leq|\rho|<1$: Very strong, $|\rho|=1$: Perfect\cite{akoglu2018user}), it does \orange{not} indicate the \textit{magnitude} or \textit{linearity} of an effect.

\begin{figure*}[h]
    \centering
    \includegraphics[width=140mm]{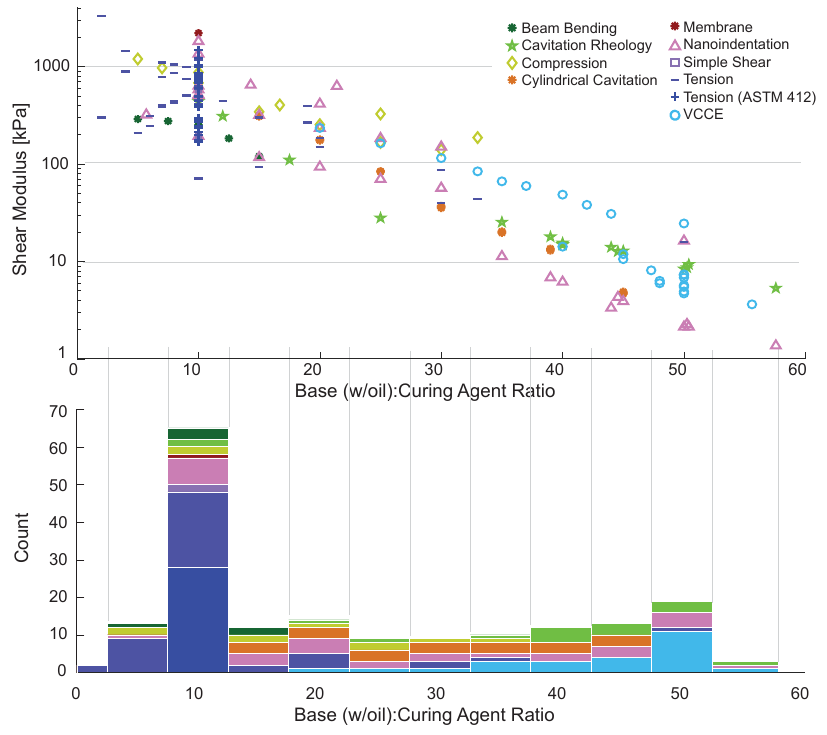}
    \caption{Stiffness data from all included studies relative to material composition show wide variability in reported moduli. Sorting by test methodology reveals trends in how test methods correspond to mix ratios. 
     The stacked histogram shows that the TDS recommended mix ratio of 10:1 wt$_{base}$:wt$_{xlinker}$ has the most reported data for moduli in literature. The abundance of data at 10:1 makes trend interpretation for combined data challenging when studies reporting only a value are averaged in with studies exploring stiffness by mix ratio. Tension testing is common for stiffer samples, however Cavity Expansion dominates in softer mix ratios while Nanoindentation and Compression have been used across a range of compositions.    }
    \label{fig:test_type_combined}
\end{figure*}

\section{Results \orange{\& Discussion}}
\label{sec:results}

\subsection{Wide modulus variability and dominance of 10:1 ratio}
Despite reportedly identical mix ratios, the measured shear moduli  vary by two orders of magnitude between different studies as shown in  Fig. \ref{fig:test_type_combined}. We observe that this is due to three main factors: 1) different test methods producing ranging results, 2) variation in specified cure schedule, and 3) the inclusion of non-reactive oils in the base mixture that are sometimes added to manipulate the viscoelasticity of the cured silicone. The histogram in Fig. \ref{fig:test_type_combined} shows a count of how many experimental results exist by test method across a range of \orange{curing agent ratios} and highlights how different test methods tend to be used for different stiffness of materials. While combining all the data is useful to inspire caution when using literature reported ``stiffness'' for SYLGARD~184, diving further into the data is necessary to try and identify the sources of the variability and the key parameters driving the reported ranges. 

Table \ref{tbl:spearman} presents the Spearman correlation between shear modulus and key process parameters: curing agent ratio, cure temperature, and cure time. Correlations are calculated for all tests and cure ratios combined \orange{using the number of data points N$_{tests}$ listed in Table \ref{tbl:spearman}}, then separately for select testing methods with data from multiple authors (and conditions) \orange{for  N$_{tests}>$10}. The second section of the table presents only tests performed at the 10:1 ratio (that recommended on the product Technical Data Sheet (TDS) \cite{dowsylgard184}),  followed by those excluding this ratio.

\orange{Tension testing results tend to be less correlated than other test methods.} All comparisons between stiffness and curing agent ratio indicate very strong correlations with an exceptionally low p-value, as expected \orange{from the reaction chemistry described in ESI.5. However, for those tests performed using tension testing, the correlation is only fair to moderate.} \orange{Examining cure temperature, when all curing agent ratios are considered, the correlation with cure temperature is higher within tension testing than across other methods. However, at the 10:1 ratio where one would expect the dependency to be most cleanly expressed given the elimination of one variable entirely, this trend does not hold and the data fail the correlation with p=0.12.} \orange{Cure time is weakly correlated when examining all tests combined. However, within individual test methods correlation routinely fails the null hypothesis and predicts a weaker or inverted correlation when evaluated with tension testing. Across all test methods and within tension testing, additional data are required to conclude that there is monotonic correlation. }

Note that in the TDS, Dow reports a tensile modulus of 6.7~MPa for a 10:1 mix ratio\cite{dowsylgard184}. Though not included in Fig. \ref{fig:test_type_combined} \orange{or Table \ref{tbl:spearman}} given \orange{incomplete information on test method and} cure conditions, this is significantly above all results compiled here.

\begin{table*}
\small
\caption{Correlations of stiffness with preparation conditions ($\rho$) and p-value of each. Very strong negative correlations between curing agent ratio and modulus across all test methods and groups of samples appears as expected. The later two sections of the table isolate samples prepared at the 10:1 w$_B$:w$_C$ in order to remove the disproportionate number of tests recorded at this ratio from biasing the results. Blanks in the table indicate cases where conditions are the same for all trials within the row.}
\label{tbl:spearman}
\begin{center}
\begin{tabular}{ l l m{1.5cm}     l m{2cm}     l m{2cm}     l l } 
\hline
Test type & N$_{articles}$ & N$_{tests}$ & \multicolumn{2}{l}{Curing agent ratio} & \multicolumn{2}{l}{Cure temperature} & \multicolumn{2}{l}{Cure time}\\
  &   &   & $\rho$ & p  & $\rho$ & p & $\rho$ & p  \\ 
\hline
All tests & 36 & 178 & -0.88 & 0.00 & 0.32 & 0.00 & -0.29 & 0.00 \\ 
\hspace{8pt}Cavitation Rheology & 3 & 16 & -0.98 & 0.00 &   &   &   &   \\ 
\hspace{8pt}Compression & 2 & 11 & -0.94 & 0.00 & 0.52 & 0.10 & -0.52 & 0.10 \\ 
\hspace{8pt}Nanoindentation & 9 & 30 & -0.90 & 0.00 & 0.72 & 0.00 & -0.37 & 0.05 \\ 
\hspace{8pt}Tension & 12 & 66 & -0.44 & 0.00 & 0.43 & 0.00 & -0.16 & 0.20 \\
\hspace{8pt}VCCE & 4 & 25 & -0.92 & 0.00 & -0.85 & 0.00 & 0.85 & 0.00 \\ 
\hline
All tests 10:1 & 27 & 53 &   &   & 0.30 & 0.03 & -0.37 & 0.01 \\
\hspace{8pt}Tension & 12 & 38 &   &   & 0.26 & 0.12 & -0.46 & 0.00 \\
\hline
All tests excluding 10:1 & 24 &  122 & -0.92 & 0.00 & 0.24 & 0.01 & -0.13 & 0.14 \\ 
\hspace{8pt}Nanoindentation & 6 & 23 & -0.91 & 0.00 & 0.61 & 0.00 & -0.05 & 0.82 \\ 
\hspace{8pt}Tension & 6 & 28 & -0.63 & 0.00 & 0.62 & 0.00 & 0.33 & 0.09 \\ 

\hline
\end{tabular}
\end{center}
\end{table*}

\subsection{False equivalencies arise due to test method}
\orange{Mix ratio and cure temperature are expected to consistently effect stiffness regardless of the test method.  
Indeed, examining Table~\ref{tbl:spearman}, correlations of stiffness with curing agent ratio and temperature are observed within a given test method. 
However, when all methods are combined this correlation is weaker.
This provides evidence of test method bias. }

\orange{ We hypothesize that the variation imparted by test method is driven by a number of factors including, but not limited to: material rate dependence, such that different methods measure the instantaneous vs relaxed modulus, the additional models required to separate stiffness from adhesion effects in contact measurements, and sample preparation and fixturing imparting imperfections in boundary conditions and defect nucleation.}

\begin{figure}[h]
    \centering
    \includegraphics[width=83mm]{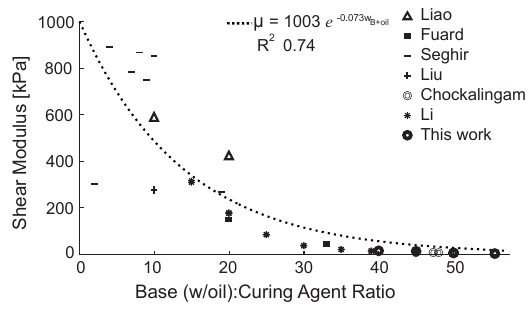}
    \caption{Test method imparts variability even when a similar cure schedule is used: 120 minutes at between 85 and 100C. Test type indicated by marker shape matching those used in Fig. \ref{fig:test_type_combined}.}
    \label{fig:all120min100C}
\end{figure}

In an effort to isolate the influence of cure schedule from that of the test method, Fig. \ref{fig:all120min100C} compares stiffness for samples cured 120 minutes at 85-100~$^\circ$C across multiple authors. 
Fitting this data with an exponential function results in $\mu=1003 e^{-0.073w_{B+oil}}$ with an R$^{2}$ of 0.74 (a linear fit on the same region had an R$^{2}$ of 0.65). Particularly for the higher curing agent ratio samples, the measured stiffness varies by hundreds of kilopascals between testing methods. Tension testing inconsistently reports moduli both higher and lower than nanoindentation, and is not used for softer samples. While the general softening trend with higher base ratios can provide a rough qualitative understanding of polymer composition's influence on stiffness, there is minimal overlap in test methods that are used between soft samples and stiff samples, leaving open the question that the trends are due to the test methodology.

\begin{figure}[h]
    \centering
    \includegraphics[width=83mm]{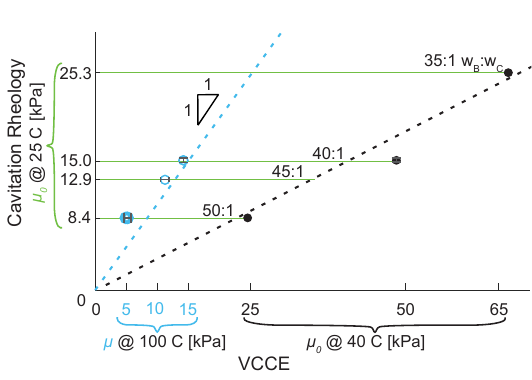}
    \caption{VCCE results of \cite{raayai2019capturing, raayai2019volume} for $\mu_{0}$ (filled marker), as well as \textcolor{cyan}{$\mu$} from Chockalingam\cite{chockalingam2021probing} and this work (\orange{unfilled \textcolor{cyan}{cyan} circle, \orange{cyan} axis labels}) compared to CR results for \textcolor{LimeGreen}{$\mu_{0}$} by Yang et al. \cite{yang2019hydraulic}\orange{. Horizontal \textcolor{LimeGreen}{green} lines specify the measured $\mu_{0}$ from cavitation rheology for comparable mix ratios (w$_B$:w$_{C}$ as noted on plot)measured with VCCE. Dashed lines are visual aids depicting both a 1:1 correlation for \textcolor{cyan}{$\mu$}:\textcolor{LimeGreen}{$\mu_0$} and lack of correlation for $\mu_0$:\textcolor{LimeGreen}{$\mu_0$}.}}
    \label{fig:VCCE vs CR}
\end{figure}

Restricting the w$_B$:w$_{C}$ ratio to a more narrow range and examining three cavity-expansion based methods, encourages caution about biases that may be more pervasive in the dataset.  Fig. \ref{fig:VCCE vs CR} compares the $\mu_0$ of Yang et al.~\cite{yang2019hydraulic} measured through CR and assuming a neo-Hookean material model, to results obtained via VCCE for comparable mix ratios (oil containing samples are excluded). 
The VCCE results from Raayai-Ardakani et al.~\cite{raayai2019capturing, raayai2019volume} are also $\mu_0$, while those of Chockalingam et al.~\cite{chockalingam2021probing} and this work are~$\mu$. 

As expected \orange{from a solid mechanics perspective}, $\mu<\mu_0$ when measured with VCCE with instantaneous moduli more than double the quasistatic moduli for comparable w$_B$:w$_{C}$ ratio. However, \orange{in these datasets, all samples used to measure $\mu$ were cured hotter and for less time than all samples to measure $\mu_0$. This results in the inversion of $\rho$ in Table \ref{tbl:spearman} for both cure temperature and time when considering VCCE independently. This inversion is another indication that the comparison of measured stiffness between different test methods and/or cure protocols can be misleading.}

\orange{Furthermore,} it can be seen on Fig.~\ref{fig:VCCE vs CR} that $\mu_0$ does not correlate 1:1 between CR and VCCE. The observed deviation between Raayai-Ardakani and Yang et al. can therefore be attributed to the 6 day room temperature cure of Yang et al. producing softer samples than the 3 days at 40~$^\circ$C cure schedule used by Raayai-Ardakani.
By contrast, the unexpected 1:1 correlation between $\mu_0$ by CR and $\mu$ by VCCE (blue line on Fig.~\ref{fig:VCCE vs CR}) is likely attributed to the discrepancy between the room temperature cure schedule used by Yang et al.~\cite{yang2019hydraulic} and the 100~$^\circ$C 2~hr schedule used by Chocaklingam et al.~\cite{chockalingam2021probing} and this work.

\subsection{Hotter cures result in stiffer material}
\begin{figure}[h]
    \centering
    \includegraphics[width=83mm]{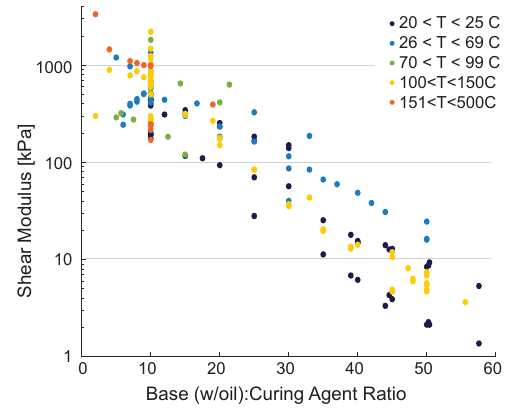}
    \caption{Cure temperatures are not equally likely across all w$_{B+oil}$:w$_C$ ratios. Room temperature cures are more common for larger ratio samples and cure temperatures over 150~$^\circ$C are rarely seen for mix ratios over 10:1.}
    \label{fig:cure_temp}
\end{figure}

An increase in cure temperature is more directly correlated to a monotonic increase measured stiffness (fair correlations considering the full data set) than is an increase cure time (weak - fair correlations). Fig. \ref{fig:cure_temp} shows that cure temperatures evaluated are not uniformly distributed amongst the base:curing agent ratios, with larger ratios being more often cured at room temperature and never over 150~$^\circ$C compared to the stiffer, low ratio, samples. This likely compounds the effect of higher mix ratio samples appearing softer due to both lower temperature and less curing agent.

Table \ref{tbl:spearman} also shows the counterintuitive result that longer cure times correlate with softer samples, however this is likely due to the significantly longer times used with room temperature curing. 
\orange{While stiffness does not follow a linear trend with all predictor variables (as shown in Fig. \ref{fig:all120min100C}) a basic regression performed on the complete data set provides insights into the relative importance of temperature, time, and curing agent ratio. Utilizing MATLAB 2023b to fit a linear regression model, cure temperature is connected to stiffness as 2.71~kPa/$^\circ$C (standard error SE of 0.74~kPa/$^\circ$C) while cure time is effected as 0.004~kPa/min cure (SE 0.006~kPa/min and a p$>$0.05, thus failing the assumption of a linear correlation). Curing agent ratio is most strongly linked to stiffness as -15.5~kPa/1 w$_{B+oil}$ (SE 1.7~kPa/1 w$_{B+oil}$). Contour plots depicting the interplay between all three predictors can be seen in Sec. 6 of the ESI.}  
With the strength of the correlation between temperature and stiffness, the tendency of lower temperature curing to produce softer samples overshadows the effect of longer cure times producing stiffer samples. 

In an effort to reduce the variability imparted by different test methods, Table \ref{tbl:spearman} is subdivided by test types and further to isolate 10:1 ratio samples. Across nanoindentation and tension testing, the correlation between cure temperature and stiffness increases to moderate - strong when test methods are considered independently, and is even stronger when 10:1 ratio samples are excluded (confirming that the number of tests performed \textit{only }at this ratio may be biasing the overall trends in results). 

There is, however, a limit at which higher temperatures will no longer produce stiffer samples and instead will degrade the polymer. Liu et al.~\cite{liu2009influences} set out to study the effect of heating on PDMS mechanical properties and also observe that stiffness increase between 100 and 150~$^\circ$C, but decreases for 200~$^\circ$C and 300~$^\circ$C samples.\footnote{The data displayed here from Liu at al.~\cite{liu2009influences} have been down selected to only the 100~$^\circ$C and 200~$^\circ$C data, though 150~$^\circ$C and 300~$^\circ$C cure temperatures were also evaluated in the original work.}

\subsection{Nonreactive oils act as part of the w$_B$ in stiffness trends}

Oil-based thinners have been proposed as a method to decouple the viscoelastic and elastic responses of PDMS without effecting the stiffness~\cite{yang2019hydraulic,mouvcka2021mechanical}. An unspecified ``thinner'' was used \orange{by Schneider et al.}~\cite{schneider2008mechanical} while \orange{Mou\v{c}ka et al.}~\cite{mouvcka2021mechanical} \orange{use 194 cSt silicone oil and Chockalingam et al.~\cite{chockalingam2021probing}, Yang et al.~ \cite{yang2019hydraulic} and the new results presented in this work use 350 cSt silicone oil added to the base}.  

\begin{figure}[h]
    \centering
    \includegraphics[width=83mm]{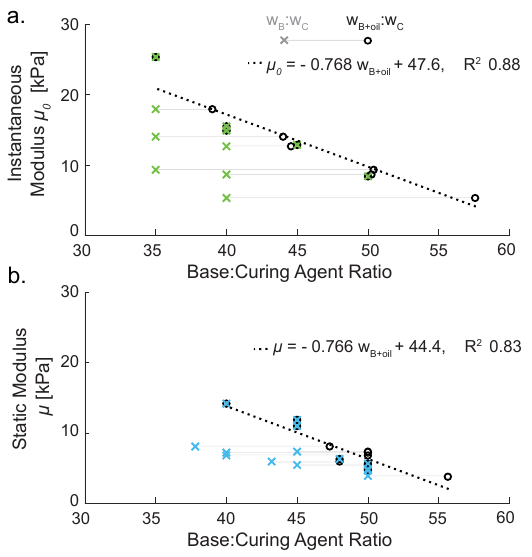}
    \caption{If w$_B$:w$_{C}$ is considered without taking oil into account (crosses), the correlation of ratio with stiffness is not as strong as when w$_{B+oil}$:w$_{C}$ is considered (circles) and the R$^2$ of a linear fit drops from 0.88 to 0.21 for CR and 0.83 to 0.31 for VCCE. The top figure shows $\mu_{0}$ from Yang et al. \cite{yang2019hydraulic} against an abscissa calculated both with and without oil included in $w_B$; the bottom plot show $\mu$ results from Chockalingam et al., \cite{chockalingam2021probing} and this work.}
    \label{fig:oil combined}
\end{figure}

Fig. \ref{fig:oil combined} shows that the correlation of shear modulus to mix compositions is stronger when the added oil is included in the $w_{B}$ than when it is not\orange{. The correlation for Yang et al. increased to} $|\rho|=0.96$ from $|\rho|=0.47$  and $|\rho|=0.85$ from $|\rho|=0.74$ combining Chockalingam et al. and this work. Across both sets of data, increasing $w_{B+oil}:w_{C}$ reduces the observed shear modulus ($\mu$ or $\mu_0$) by a similar amount (0.768 vs 0.766 kPa per added $w_{B+oil}$ with a linear fit to the data in this region). While all authors report changes to the viscoelastic time constants when including nonreactive oils, the \textit{trend} in softening with increasing $w_{B+oil}$ for soft PDMS is shown to be consistent a) regardless as to whether the ratio is altered with base or oil, and b) in both instantaneous and quasistatic modulus.

\subsection{\orange{Reaction chemistry offers hypotheses for variation}}

A number of the studies point out changes in mechanical properties of SYLGARD 184 within days to weeks after completing cure \cite{chockalingam2021probing, cheng2011characterization}. Cheng et al. showed nearly a doubling in stiffness with 18 vs 3 days of storage at room temperature after an initial cure. 
\orange{In their work with pure PDMS, Esteves et al. propose that secondary reactions occur as PDMS polymerizes \cite{esteves2009influence}. They observed that consumption of Si-H groups in the cross linker  cannot be fully explained by hydrosylation (the conventionally understood pathway of PDMS cross linking) thus requiring that additional pathways are present. These additional pathways would be effected by ambient oxygen and moisture, pointing to relative humidity during fabrication as an additional parameter to consider in fabrication and storage of PDMS. }

\orange{Recent work by Bardelli et al.~\cite{bardelli2022influence} points to cure temperature as a factor in changing the molecular weight between crosslinks $M_c$ in PDMS. As described in ESI.5, $M_c$ is directly connected to the stiffness of a polymer. Bardelli et al. observe nearly a doubling of $\mu$ between a 4 hour cure at 65 $^{\circ}$C compared to 150 $^{\circ}$C (with $M_c$ decreasing from 2050 to 1050 g/mol).}
\orange{Additionally, if lower temperature cures allow for more secondary reactions to take place, this could explain the resulting solid being softer. }

\section{Conclusions}
\label{sec:conclusion}
\orange{The choice of mechanical testing method can have a significant influence of the measured stiffness of soft materials. 
Using commercially available PDMS (SYLGARD 184) as a test case, we examine this bias by combining new experimental data and that  surveyed from 32 published papers, spanning 9 different testing methods. Statistical analysis  reveals that bias imparted by test method is compounded by inconsistent curing approaches. The current methods of both testing and reporting are thus far from providing reproducible results for the mechanical properties of SYLGARD~184 and, by extension, for other soft and biological materials. A number of steps must be taken to better observe the properties that so many authors to date have claimed to measure. } 

\orange{Test method has been implicated previously as returning variable measurements within identical soft material samples, and PDMS is no exception.} Tension tests on PDMS reported the widest variation of stiffness, with nanoindentation also providing widely variable results. Though these are both common testing techniques, we recommend caution in the use of both methods. At low mixing ratios PDMS is brittle~\cite{seghir2015extended} and tension testing is highly sensitive to surface imperfections on the sample, while at higher w$_B$:w$_C$ ratios, the soft modulus means that results are influenced by the clamping conditions, \orange{without a guarantee that it is possible to set up a successful test at all}. The variability of nanoindentation is likely due to the uncertain nature of the adhesion between the sample and indenter tip, which was the focus of many of the authors referenced here. Using such a method that relies on consistent surface adhesion, in a material where surface treatments are applied to widely varying degrees, will continue to produce imprecise results.

\orange{The limitations of testing make it challenging to characterize the  composition-property relationship of SYLGARD 184. This difficulty is further accentuated by the inconsistency of fabrication conditions and reporting of curing protocols.} At present, cure schedules ranged widely between researchers with very little overlap if the authors were not currently or historically from the same research group. \orange{Of utmost importance is fully reporting \orange{the cure schedule and preparation method}, as well as timing between cure and test. Unifying how PDMS is cured to obtain well characterized responses could help to enhance data comparisons and reliability. Hence, we would propose three temperatures for PDMS where control of mechanical properties is desired: a $25^\circ$C room temperature, a ``moderate'' $60 ^\circ$C at or below the glass transition temperature of plastics commonly used in additive manufacturing, and a ``hot'' $100^\circ$C for rapid curing. Curing times may vary depending on sample geometry. }

Each researcher has unique needs and capabilities for their research, and not each mechanical testing method is possible for a given sample. As soft materials research continues to expand into more complex and biological systems that present additional challenges for reproducible measurements, building agreement around common synthetic soft materials is worthwhile for the community. By noting the current discrepancies in literature on PDMS and by advocating that work going forward place a greater emphasis on consistency and comparison to the full body of literature (not selected works that happen to align with ones results), we anticipate better agreement in measured mechanical properties is easily attainable and hope that this manuscript motivates this as a community goal.

\section*{Author Contributions}
\textbf{Hannah Varner}: Writing - Original Draft, Writing - Review \& Editing, Conceptualization, Investigation, Data Curation, Methodology, Formal Analysis, Visualization. \textbf{Tal Cohen}: Conceptualization, Writing - Review \& Editing, Resources, Supervision.

\section*{Conflicts of interest}
There are no conflicts to declare.

\section*{Acknowledgements}
We acknowledge the partial support of our work through Office of Naval Research grant N000142312530, and the support from the National Science Foundation under award number CMMI-1942016..
H.V. acknowledges the support of the Department of Defense  through the National Defense Science \& Engineering Graduate Fellowship Program. 
Funding agencies were not involved in study design; in the collection, analysis and interpretation of data; in the writing of the report; or in the decision to submit the article for publication.


\balance

\bibliography{rsc} 
\bibliographystyle{rsc} 

\end{document}